\documentclass{emulateapj}
\usepackage{apjfonts}
\usepackage{graphicx}
\usepackage{subfigure}
\usepackage{placeins}
\newcommand{\secpoint}{\mbox{$''\mskip-7.6mu.\,$}}

\slugcomment{Accepted for publication in the Astrophysical Journal, Apr 6 (2014)}

\begin{document}

\title{Gemini Long-slit Observations of Luminous Obscured Quasars: Further Evidence for an Upper Limit on the Size of the Narrow-Line Region}

\shorttitle{Gemini Spectroscopy of Obscured Quasars}
\shortauthors{HAINLINE ET AL.}

\author{\sc Kevin N. Hainline, Ryan C. Hickox}
\affil{Department of Physics and Astronomy, Dartmouth College, Hanover, NH 03755}

\author{\sc Jenny E. Greene}
\affil{Department of Astrophysical Sciences, Princeton University, Princeton, NJ 08544}

\author{\sc Adam D. Myers}
\affil{Department of Physics and Astronomy, University of Wyoming, Laramie, WY 82071}

\author{\sc Nadia L. Zakamska, Guilin Liu}
\affil{Department of Physics and Astronomy, The Johns Hopkins University, Baltimore, MD 21218}

\author{\sc Xin Liu\altaffilmark{1}}
\affil{Department of Physics and Astronomy, University of California, Los Angeles, CA 90095}

\author{}

\altaffiltext{1}{Hubble Fellow}

\begin{abstract}

We examine the spatial extent of the narrow-line regions (NLRs) of a sample of 30 luminous obscured quasars at $0.4 < z < 0.7$ observed with spatially resolved Gemini-N GMOS long-slit spectroscopy. Using the [OIII]$\lambda5007$ emission feature, we estimate the size of the NLR using a cosmology-independent measurement: the radius where the surface brightness falls to 10$^{-15}$ erg s$^{-1}$ cm$^{-2}$ arcsec$^{-2}$. We then explore the effects of atmospheric seeing on NLR size measurements and conclude that direct measurements of the NLR size from observed profiles are too large by 0.1 - 0.2 dex on average, as compared to measurements made to best-fit S\'{e}rsic or Voigt profiles convolved with the seeing. These data, which span a full order of magnitude in IR luminosity ($\log{(L_{8 \mu \mathrm{m}} / \mathrm{erg\, s}^{-1})} = 44.4 - 45.4$) also provide strong evidence that there is a flattening of the relationship between NLR size and AGN luminosity at a seeing-corrected size of $\sim 7$ kpc. The objects in this sample have high luminosities which place them in a previously under-explored portion of the size-luminosity relationship. These results support the existence of a maximal size of the narrow-line region around luminous quasars; beyond this size either there is not enough gas, or the gas is over-ionized and does not produce enough [OIII]$\lambda5007$ emission.
\end{abstract}


\section{Introduction}
\label{sec:intro}

\begin{deluxetable*}{lccccr}
\tabletypesize{\scriptsize}
\tablecaption{Sample and Observations \label{tab:sample}}
\tablewidth{0pt}
\tablehead{
\colhead{SDSS Name} & \colhead{$z$} & \colhead{Obs. Date} & \colhead{Seeing} & \colhead{Exp. Time} & \colhead{PA\tablenotemark{a}} \\
\colhead{} & \colhead{} & \colhead{} & \colhead{(arcsec)} & \colhead{(s)} & \colhead{}
}
\startdata

J005621.72+003235.7 & 0.484 & 2007 Jan 27 & 0.63 & 5400 & 345 \\
J013416.34+001413.6 & 0.557 & 2006 Dec 25 & 0.52 & 3600 & 90 \\
J015716.92--005304.6 & 0.422 & 2007 Jan 01 & 0.56 & 5400 & 100 \\
J020655.71+010826.6 & 0.471 & 2009 Aug 22 & 0.51 & 3600 & 352 \\
J021047.00--100152.9 & 0.540 & 2007 Jan 25 & 0.58 & 3600 & 180 \\
J030425.70+000740.8 & 0.556 & 2009 Sep 25 & 0.29 & 3600 & 75 \\
J031449.09--010502.2 & 0.558 & 2009 Sep 25 & 0.48 & 1800 & 228 \\
J031909.61--001916.6 & 0.635 & 2009 Sep 25 & 0.38 & 1800 & 206 \\
J031950.52--005850.4 & 0.626 & 2006 Dec 27 & 0.36 & 3600 & 130 \\
J075944.64+133945.5 & 0.649 & 2009 Oct 21 & 0.95 & 3600 & 212 \\
J080154.25+441233.9 & 0.556 & 2006 Dec 29 & 0.93 & 3600 & 0 \\
J080754.51+494627.5 & 0.575 & 2009 Nov 11 & 0.93 & 3369 & 309 \\
J081404.55+060238.3 & 0.561 & 2009 Nov 20 & 0.59 & 3600 & 45 \\
J081507.41+430426.9 & 0.510 & 2009 Nov 20 & 0.95 & 3600 & 50 \\
J082313.50+313203.7 & 0.433 & 2006 Dec 29 & 0.80 & 3600 & 115 \\
J083134.21+290239.4 & 0.568 & 2009 Dec 15 & 1.04 & 1800 & 214 \\
J084339.47+290124.5 & 0.686 & 2009 Nov 20 & 0.71 & 3600 & 130 \\
J085231.35+074013.4 & 0.420 & 2009 Nov 22 & 0.68 & 3600 & 87 \\
J091819.66+235736.4 & 0.419 & 2009 Nov 22 & 0.62 & 3600 & 154 \\
J092152.46+515348.0 & 0.588 & 2009 Dec 12 & 0.54 & 3600 & 8 \\
J094312.81+024325.8 & 0.592 & 2009 Dec 14 & 0.63 & 4320 & 346 \\
J094311.57+345615.8 & 0.530 & 2006 Dec 31 & 0.71 & 3600 & 115 \\
J095019.90+051140.9 & 0.524 & 2006 Dec 31 & 0.70 & 3600 & 210 \\
J101322.12+272209.4 & 0.666 & 2009 Dec 15 & 0.87 & 3600 & 238 \\
J102746.04+003205.0 & 0.614 & 2009 Dec 16 & 0.82 & 3600 & 220 \\
J103822.08+523115.8 & 0.599 & 2009 Dec 11 & 0.54 & 3600 & 322 \\
J104210.19+382255.3 & 0.608 & 2009 Dec 16 & 1.02 & 3600 & 20 \\
J104402.39+300834.0 & 0.497 & 2009 Dec 25 & 0.85 & 3600 & 286 \\
J104731.84+063603.7 & 0.435 & 2009 Dec 25 & 0.55 & 3600 & 42 \\
J105056.15+343703.3 & 0.491 & 2009 Dec 18 & 0.78 & 3600 & 27 \\

\enddata
\tablenotetext{a}{In degrees east of north.}
\end{deluxetable*}

	Current theories of galaxy formation and evolution imply a role for Active Galactic Nuclei (AGNs) which are powered by accretion from a supermassive black hole. The intense radiation from the disk around an accreting supermassive black hole will ionize the surrounding gas, producing characteristic emission features in an AGN spectrum that can be measured in spatially resolved images and spectroscopy, even out at large galactocentric distances \citep{boronson1984, stocktonmackenty1987}. Early research uncovered regions of ionized gas at kpc scales around radio-loud quasars \citep{wampler1975, stockton1976}, and more recently, the size of extended narrow-line emission has been explored as a function of AGN luminosity. \citet{bennert2002} used Hubble Space Telescope (HST) imaging for a sample of radio-quiet quasars to describe a relationship between the size of the AGN narrow-line region (NLR) (as traced by the narrow emission line [OIII]$\lambda$5007) and AGN luminosity. This work was later confirmed by \citet{schmitt2003} using HST imaging of a larger sample of Seyfert 1 and Seyfert 2 galaxies at lower luminosities, and then at high luminosities in both \citet{greene2011}, \citet{liu2013}, and \citet{hainline2013} who examined samples of Type II quasars using both long-slit and IFU spectroscopy to carefully probe NLR emission. 
	
	Historically, [OIII]$\lambda$5007 emission line strength has been used as both a proxy for the size of the NLR and a measure of the intrinsic AGN luminosity. In \citet[][hereafter H13]{hainline2013}, the authors compared NLR size to AGN luminosity as derived from mid-IR photometry gathered by the Wide Field Infrared Explorer \citep[WISE,][]{wright2010}. For powerful AGNs, the mid-IR traces emission from warm to hot dust near the central supermassive black hole \citep{pier1993}, and AGN IR luminosity has been shown to correlate with AGN soft X-ray luminosity \citep[e.g.][]{krabbe2001, lutz2004, horst2008, asmus2011, matsuta2012}. Importantly, as IR luminosity does not depend on properties of the NLR, the relationship between the NLR size and the AGN luminosity as traced by IR emission provides important insights into both the mechanism and the physical extent of the power of a given AGN. H13 report a steep trend to this relationship at the low-luminosity end, as well as a possible flattening at the high luminosity end, which they claim results from the most powerful AGNs effectively ionizing all of the available gas above a given density. Evidence for this flattening, however, was primarily based on a small sample of IR-luminous AGNs from the literature. 
		
	In this paper, we improve our measurement of the high-luminosity end of the NLR-size-L$_{\mathrm{AGN}}$ relationship, examining a sample of 30 luminous Type II quasars from the \citet{zakamska2003} and \citet{reyes2008} SDSS-selected sample using long-slit spectroscopy with the GMOS instrument on Gemini-N, following the method presented in \citet{greene2011} and H13. This sample spans a wide range in IR luminosities, and by measuring the spatial extent of the NLR in these objects, we can better describe the flattening of the relationship between IR luminosity and NLR size, particularly at the high-luminosity end. Importantly, we discuss how measurements of the size of the NLR can be affected by atmospheric seeing, and introduce a surface brightness profile modeling procedure to account for seeing effects. 
		
	We describe our Type II quasar sample and Gemini GMOS observations in Section \ref{sec:sample}, discuss the measurements of the spatial sizes of the observed NLRs in Section \ref{sec:nlrsizes}, examine the relationship between NLR size and AGN IR luminosity in Section \ref{sec:nlrsize_relationship}, and discuss the excitation properties of the quasar NLR observations in Section \ref{sec:excitation}. Finally, we present our conclusions in Section \ref{sec:conclusions}. Throughout, we assume a standard $\Lambda$CDM cosmological model with $H_0 = 71$ km s$^{-1}$ Mpc$^{-1}$, $\Omega_{M} = 0.27$, and $\Omega_{\Lambda} = 0.73$ \citep{komatsu2011}.
	
\section{Quasar Sample, Observations and Data Reduction}
\label{sec:sample}

	\begin{figure*}[]
	\epsscale{1.05} 
	\plotone{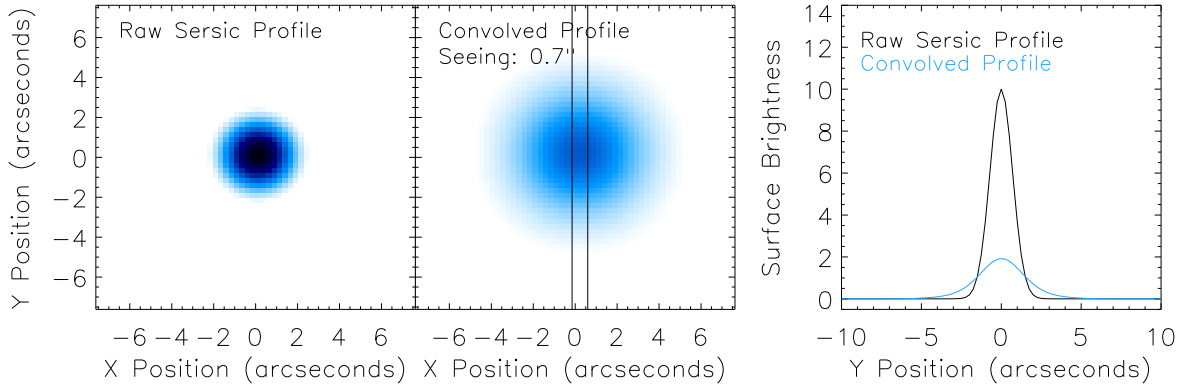}
	\caption{
	\label{fig:MethodCompare} Our modeling procedure starts with a two-dimensional surface brightness profile (left), which is then convolved with a two-dimensional Moffat profile matched to the seeing (in this example, the FWHM of the atmospheric seeing is $0.7"$) to produce a convolved profile (middle). In the middle panel, we also overplot the $0.75"$ slit used in the bulk of our observations. To match the two-dimensional profile with the observed one-dimensional profiles, we averaged the convolved profiles across the slit, and plot both the raw one-dimensional S\'{e}rsic profile as well as the convolved profile together with arbitrary surface brightness units (right).} 
	\epsscale{1.}
         \end{figure*}
         
	The sample of objects examined in this work were selected from the SDSS Type II quasar sample first described in \citet{zakamska2003} and \citet{reyes2008}. From this sample, we chose 30 luminous quasars with redshifts in the range $0.4 < z < 0.7$ with $\log (L_{[\mathrm{OIII}]\lambda5007}/L_{\odot})>9.4$ in order to maximize the likelihood of detecting spatially resolved NLR emission at large galactic radii. Five objects were selected that overlap with the sample from \citet{liu2013}: J0210-1001, J0319-0019, J0319-0058, J0759+1339, and J0807+4946. We use these objects to compare our size measurements with previously published results. Finally we used observations at 1.4 GHz from the FIRST survey \citep{becker1995,white1997} to determine that all but two objects (J0843+2901 and J0943+0243) are radio-quiet based on their position on the $L_{\mathrm{[OIII]}}$ versus $\nu L_\nu$ (1.4 GHz) diagram \citep{xu1999,zakamska2004}. We will include these objects in our analysis, but mark them separately in our plots.
	
	The full sample of objects was observed with the GMOS instrument on Gemini-N in two queued campaigns, one in 2006 (GN-2006B-Q-101, PI: N. Zakamska), and the other in 2009 (GN-2009B-Q-55, PI: X. Liu). For both campaigns, the objects were observed in GMOS slit mode with the R400-G5305 grating (at a resolution of $R \sim 1900$) and an observed wavelength range of $5000 - 8000$ \AA. At the average redshift of our sample, this range corresponds to rest-frame wavelengths $3200 - 5200$ \AA. Each object was observed for a minimum of 3600s (except for J0807+4946, which was observed for only 3369s, and J0314--0105, J0319--0019, and J0831+2902 which were only observed for 1800s). For the 2006 observations, a slit of width $0.5''$ was used, while for the 2009 observations, a slit of width $0.75''$ was used. At the average redshift of the sample, $1''$ corresponds to 6.3 kpc. For each observation, the slit was first centered on the quasar, and then oriented to cover a nearby neighboring object. For these observations, the average and median seeing, as measured from acquisition images taken before the spectra, was a Moffat FWHM of $0.7''$, with a range between $0\secpoint3$ and $1\secpoint0$. We spatially resolve all but five of the quasar NLRs in our sample, and report upper limits on the NLR sizes for the unresolved objects. The full sample of objects, including redshifts, observation dates, exposure times, seeing, and position angles is described in Table \ref{tab:sample}. We also observed two white dwarfs as flux standards, G191B2B and EG131. 
	
	The data were reduced in IRAF\footnote{IRAF is distributed by the National Optical Astronomy Observatory, which
is operated by the Association of Universities for Research in Astronomy, Inc., under cooperative agreement with the National Science Foundation.} using the Gemini GMOS reduction package, in a manner similar to the analysis of \citet{liu2009}. The primary steps that were undertaken on the data were bias subtraction, flat fielding, interpolation across the chip gaps, cosmic ray and bad pixel cleaning, and wavelength calibration using arc exposures taken before and after each observation. The end product was a fully reduced, two-dimensional spectrum from which a spatial profile of the [OIII]$\lambda$5007 could be extracted.

	For all but one of the objects in our sample, we found the corresponding WISE source from the WISE All-Sky Source Catalog. At the redshift range of the sample, we estimate the rest-frame IR luminosity using the WISE [4.6], [12], and [22] bands, which we interpolated in log-log space to estimate the flux and luminosity at rest-frame 8$\mu$m ($L_{8 \mu \mathrm{m}}$) for each object. Similar to the procedure in H13, we model the AGN mid-IR emission with a power law and do not account for the individual filter response functions, although, based on the WISE colors for these objects, any flux corrections would be on the order of a few percent \citep{wright2010}. We also assume that the flux is measured at the central wavelength for each filter. For the quasar J0157--0053, the WISE photometry is contaminated by the presence of a nearby bright star, and so we do not report 8$\mu$m luminosities for this object.

\section{Narrow-Line Region Sizes}
\label{sec:nlrsizes}

	We can use the spatially resolved GMOS spectra to measure the spatial extent of the [OIII]$\lambda$5007 emission line feature, which provides us a measure of the NLR size that we can compare to other studies. We first collapsed each two-dimensional spectrum in the wavelength direction over a wavelength range with a width twice the FWHM of [OIII]. We used the same procedure on the corresponding spectroscopic standard two-dimensional spectrum, measured the total flux in this region (corrected for slit losses), and, along with a flux calibrated reference spectrum for the standard star, we estimated a flux correction factor which we applied to the quasar spatial profiles to convert from the observed units to erg s$^{-1}$ cm$^{-2}$ arcsec$^{-2}$. As in H13, in order to properly compare the extent of the [OIII] emission line with those measurements from other authors, we follow the prescription from \citet{liu2013} and parameterize the NLR size with $R_{\mathrm{int}}$, calculated as the size of the NLR at a limiting surface brightness corrected for cosmological dimming of 10$^{-15}$ erg s$^{-1}$ cm$^{-2}$ arcsec$^{-2}$. As this measure of the NLR size is not dependent on the depth of the observations, it is ideal for comparing size measurements made with different instruments and exposure times.

	\begin{figure*}[!ht]
	\epsscale{0.9} 
	\plotone{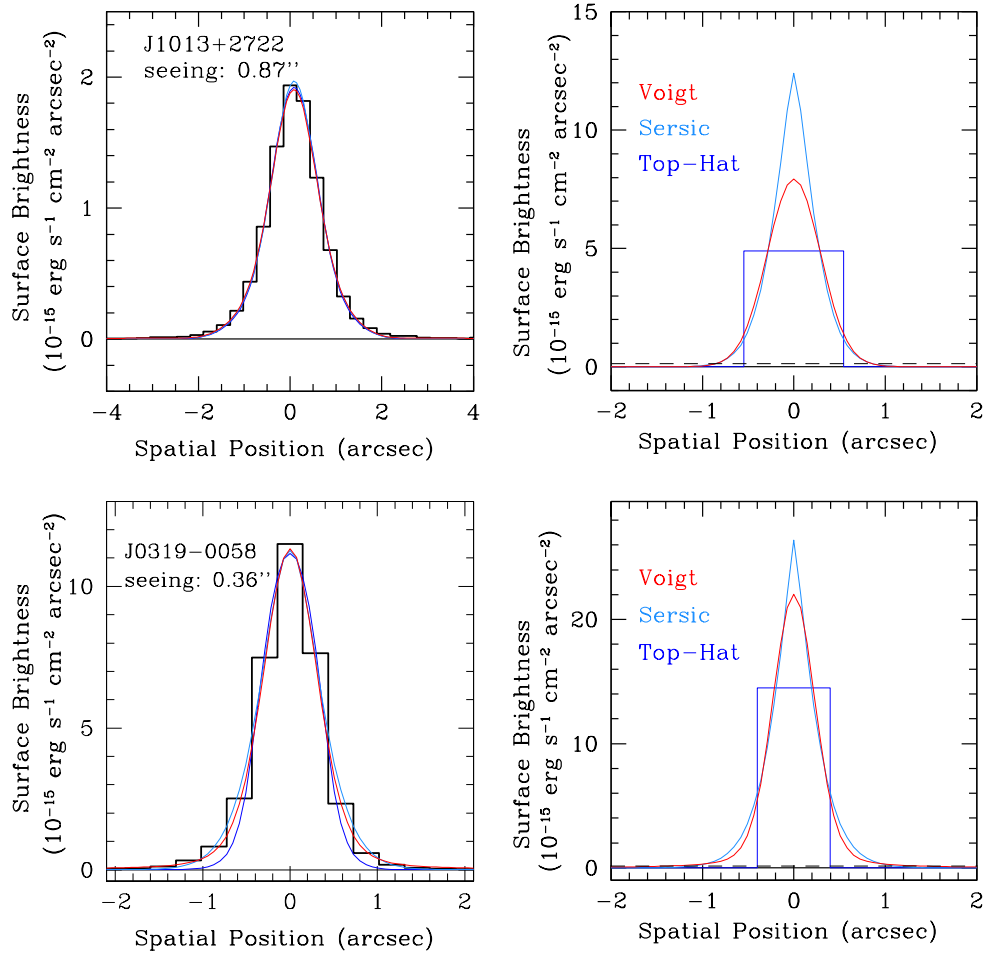}
	\caption{
	\label{fig:ProfileCompare} Example surface brightness profile for the quasars J1013+2722 ($z = 0.666$, top) and J0319-0058 ($z = 0.626$, bottom). The left plots show the observed surface brightness profiles in black, overplotted with the best-fitting models convolved with a Moffat profile scaled to match the seeing. The right plots show the intrinsic top-hat (blue), S\'{e}rsic (light blue), and Voigt (red) models, along with the surface brightness limit down to which $R_{\mathrm{int}}$ was measured (dashed line). For J1013+2722, the top-hat model results in a smaller NLR size than the S\'{e}rsic or Voigt profile models, while the effect of the seeing is to produce similar observed profiles. For J0319-0058, which represents the majority of the objects in our sample, the best-fit top-hat model does not account for the extended wings, leading to an underpredicted NLR size.} 
	\epsscale{1.}
         \end{figure*}

	Ground-based measurements of spatial profile sizes can be affected by atmospheric seeing, which, even in spatially resolved observations, will artificially increase observed NLR sizes. The seeing, then, must be carefully accounted for in order to compare sizes measured under different observing conditions. Previous studies of the NLR size, including \citet{greene2011} and \citet{liu2013}, provide NLR size estimates measured directly from the observed spatial profile where the observations resolve the NLR above the measurement of the seeing. Since these measurements are not corrected for the seeing, they represent an upper limit on the potential size of the NLRs for a given quasar. For the objects in our sample, we perform an equivalent measurement on the observed spatial profiles. The NLR sizes we measure in this way span a similar range compared to those reported for similar objects in \citet{greene2011} and \citet{liu2013}, with an average (median) log($R_{\mathrm{int}}$ / pc)$ = 4.1 (4.0) \pm 0.1$. 
	
	Here, we use our GMOS spectra to carefully examine the effect that seeing can have on NLR size measurements and demonstrate the size discrepancy between measurements made directly from the observed spatial profile and those from models for the intrinsic spatial profile. In H13, the authors modeled the intrinsic one dimensional surface brightness of each object with a S\'{e}rsic profile convolved with a Moffat profile scaled to match the seeing for the individual observation. A Moffat model of the seeing was chosen as this profile has been shown to robustly model astronomical point-spread functions \citep[PSFs, e.g.][]{trujillo2001}. 
	
	As the Gemini observations were taken under much better seeing conditions than the SALT observations in H13, we have expanded this procedure in a few key ways. First, instead of performing the convolution with the seeing in one dimension, for each quasar in our sample we created a two-dimensional profile with units of surface brightness. These two-dimensional profiles were modeled with finer precision than the actual pixel scale of the observations. We then convolved the intrinsic profile with two dimensional Moffat profiles matched to the seeing to replicate the observed profile. From here, we averaged across the slit to create one-dimensional profiles in units of surface brightness that we compared to the observed profiles. This process is illustrated in Figure \ref{fig:MethodCompare}, where we show a two-dimensional intrinsic profile on the left, the convolved profile (with our average seeing of $0.7''$) in the middle, and then the corresponding one-dimensional surface brightness profiles on the right.
	 
	 For this procedure, we assumed that the true surface brightness profile is circularly symmetric on the sky, which is supported by resolved IFU observations of these objects given in \citet{liu2013}. As we do not actually know the form of the intrinsic NLR surface brightness profiles, we chose to model the NLRs with three different analytic functions: a ``top-hat'' profile, a S\'{e}rsic profile and a Voigt profile. The top-hat profile assumes that the NLR occupies a circular region of constant surface brightness, and at a specified radius the profile drops to 0. While this is an unphysical model, for each object, it provides a lower limit on the NLR size for a given observed, resolved profile. It is important to note that for the majority of the objects, a top-hat profile convolved with the seeing provides a poor fit to the data, as the seeing alone cannot account for the extended wings of the observed surface brightness profiles. For this reason, we also model each object with both a S\'{e}rsic profile and a Voigt profile. Both the Voigt and S\'{e}rsic profiles provide equally good fits to the data, based on the reduced $\chi^2$ values we measured. From each fit, we then calculated $R_{\mathrm{int}}$ using the intrinsic surface brightness profile, before it is smeared out by the seeing. 

\begin{deluxetable*}{lcccccc}
\tabletypesize{\scriptsize}
\tablecaption{AGN Luminosity and NLR Sizes \label{tab:nlrsizes}}
\tablewidth{0pt}
\tablehead{
\colhead{SDSS Name} & \colhead{$z$} & \colhead{log($L_{8\mu \mathrm{m}}$)} & \colhead{log($R_{\mathrm{int}}$ / pc)} 
& \colhead{log($R_{\mathrm{int}}$ / pc)} & \colhead{log($R_{\mathrm{int}}$ / pc)} & \colhead{log($R_{\mathrm{int}}$ / pc)} \\
\colhead{} & \colhead{} & \colhead{erg s$^{-1}$} & \colhead{No Fit} & \colhead{Top-Hat} & \colhead{Voigt} & \colhead{S\'{e}rsic}
}
\startdata

J0056+0032 & 0.484 & 45.09 & 3.91 & 3.47 & 3.80 & 3.76 \\
J0134+0014 & 0.557 & 45.02 & 3.96 & 3.50 & 3.95 & 3.79 \\
J0157--0053 & 0.422 & -\tablenotemark{a} & 3.95 & 3.38 & 3.91 & 3.85 \\
J0206+0108 & 0.471 & 44.65 & 3.95 & 3.33 & 3.90 & 3.75 \\
J0210--1001 & 0.540 & 44.98 & 4.09 & 3.41 & 3.72 & 3.65 \\
J0304+0007 & 0.556 & 45.00 & 3.84 & 3.42 & 3.75 & 3.74 \\
J0314--0105 & 0.558 & 44.78 & 3.99 & 3.42 & 3.79 & 3.71 \\
J0319--0019 & 0.635 & 44.94 & 3.98 & 3.48 & 3.94 & 3.81 \\
J0319--0058 & 0.626 & 44.80 & 3.96 & 3.44 & 3.97 & 3.87 \\
J0759+1339 & 0.649 & 45.28 & $<$4.10 & - & - & - \\
J0801+4412 & 0.556 & 44.93 & $<$3.95 & - & - & - \\ 
J0807+4946 & 0.575 & 45.20 & 4.20 & 3.60 & 4.06 & 4.01 \\
J0814+0602 & 0.561 & 44.93 & 4.07 & 3.57 & 4.04 & 3.97 \\
J0815+4304 & 0.510 & 45.39 & 4.26 & 3.61 & 4.05 & 3.95 \\
J0823+3132 & 0.433 & 44.80 & 4.08 & 3.65 & 3.95 & 3.93 \\
J0831+2902 & 0.568 & 44.71 & 4.15 & 3.58 & 4.01 & 3.91 \\
J0843+2901 & 0.686 & 45.01 & 3.97 & 3.62 & 3.81 & 3.82 \\
J0852+0740 & 0.420 & 45.28 & 4.02 & 3.33 & 3.69 & 3.70 \\
J0918+2357 & 0.419 & 44.64 & $<$3.89 & - & - & - \\
J0921+5153 & 0.588 & 45.06 & 3.92 & 3.40 & 3.64 & 3.64 \\
J0943+0243 & 0.592 & 44.36 & 3.95 & 3.49 & 3.80 & 3.74 \\ 
J0943+3456 & 0.530 & 45.20 & 4.06 & 3.30 & 3.80 & 3.58 \\
J0950+0511 & 0.524 & 45.11 & 4.04 & 3.56 & 3.97 & 3.86 \\
J1013+2722 & 0.666 & 45.59 & 4.01 & 3.56 & 3.77 & 3.78 \\
J1027+0032 & 0.614 & 44.98 & 4.08 & 3.52 & 3.86 & 3.79 \\
J1038+5231 & 0.599 & 44.97 & 4.09 & 3.49 & 4.10 & 3.90 \\
J1042+3822 & 0.608 & 45.35 & $<$3.92 & - & - & - \\
J1044+3008 & 0.497 & 44.80 & $<$4.05 & - & - & - \\
J1047+0636 & 0.435 & 44.76 & 3.91 & 3.44 & 3.76 & 3.75 \\
J1050+3437 & 0.491 & 44.84 & 3.99 & 3.57 & 3.85 & 3.85 \\
\enddata
\tablenotetext{a}{The WISE photometry for this object is contaminated with that from a nearby star.}
\end{deluxetable*}

	We show the surface brightness profile for two representative objects, J1013+2722 ($z = 0.666$, seeing $= 0.87''$), and J0319--0058 ($z = 0.626$, seeing $= 0.36''$), in Figure \ref{fig:ProfileCompare}. In this figure, the left plots show the observed profile (log($R_{\mathrm{int}}$ / pc)$ = 4.01$ for J1013+2722, and log($R_{\mathrm{int}}$ / pc)$ = 4.0$ for J0319--0058) with the best-fit models overplotted, and the corresponding intrinsic profiles are shown in the right plots. On this figure, we also show the surface brightness limits (corrected for cosmological dimming) used to measure $R_{\mathrm{int}}$ with a dashed line. For J1013+2722, all three fit the data equally well when the seeing is applied, but while the Voigt and S\'{e}rsic profiles result in sizes that are very similar (log($R_{\mathrm{int}}$ / pc)$ \sim 3.8$), a top-hat profile results in sizes that are 0.2 dex smaller (log($R_{\mathrm{int}}$ / pc)$ = 3.6$). For J0319--0058, while the convolved Voigt and S\'{e}rsic profiles fit the data well, the top-hat profile does not fit the wings of the observation, leading to an artificially small intrinsic profile.

	\begin{figure*}[ht]
	\epsscale{0.9} 
	\plotone{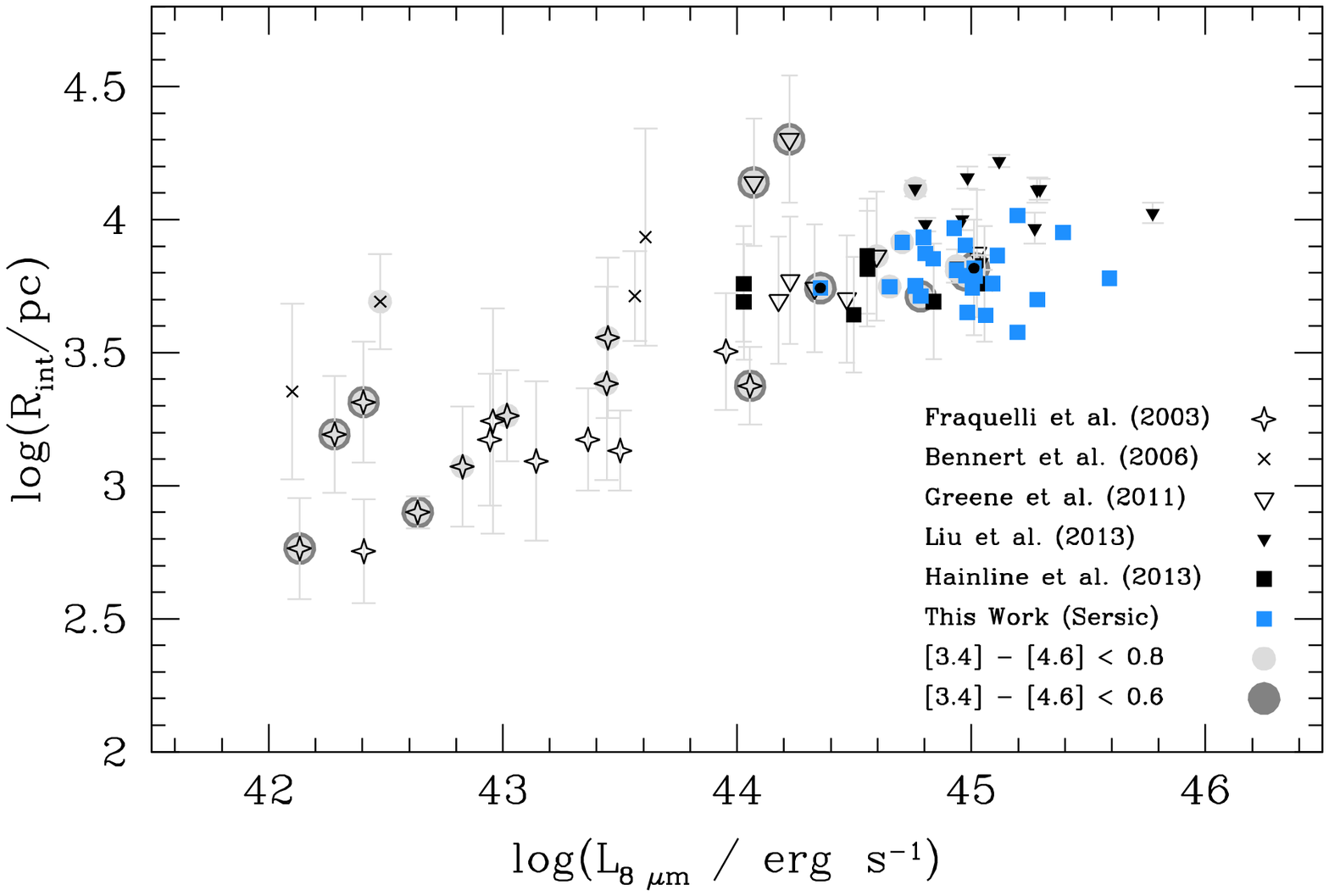}
	\caption{
	\label{fig:NLRsize} Radii of the [OIII]$\lambda$5007-emitting region plotted against the AGN IR luminosity. The NLR size is represented by $R_{\mathrm{int}}$, which is defined as the size of the object at a limiting surface brightness corrected for cosmological dimming of $10^{-15}/(1+z)^4$ erg s$^{-1}$ cm$^{-2}$ arcsec$^{-2}$ \citep{liu2013}. We plot $R_{\mathrm{int}}$ measured using a S\'{e}rsic profile for the Type II quasars from our GMOS sample using light blue squares. We also plot multiple samples from the literature, including Seyfert galaxies from \citet[][open stars]{fraquelli2003} and \citet[][Xs]{bennert2006}, as well as obscured quasars from \citet[][open triangles]{greene2011}, \citet[][black filled triangles]{liu2013}, and \citet[][black squares]{hainline2013}. We plot those objects with WISE color $[3.4]-[4.6] < 0.8$ superimposed on a light gray circle, and those objects with WISE color $[3.4]-[4.6] < 0.6$ are plotted over a dark gray circle, to indicate which objects may be suffering from contamination by stellar processes. We indicate the two radio-loud objects in our sample, J0843+2901 and J0943+0243, with black circles. The Gemini sample from this work provides more evidence for the flattening of the relationship between NLR size and AGN luminosity at the luminous end.} 
	\epsscale{1.}
         \end{figure*}

	We report the calculated sizes for the objects in our sample in Table \ref{tab:nlrsizes}. In this Table, we give our estimated values of $L_{8\mu \mathrm{m}}$ along with the sizes measured directly from the observed profiles as well as the sizes derived from fitting with a top-hat profile, a Voigt profile, and a S\'{e}rsic profile. For those objects where we do not resolve the spatial profile above the seeing, we only report the upper limits derived from directly measuring $R_{\mathrm{int}}$ without fitting the data. It can be seen from Table \ref{tab:nlrsizes} that the sizes measured using a Voigt profile are similar, but slightly larger (by 0.07 dex on average) than those measured using a S\'{e}rsic profile. In H13, the authors find that S\'{e}rsic profiles result in significantly smaller sizes than those measured from Voigt profiles, but this is due to the much larger seeing ($\sim 2''$) in their data. Most importantly, the difference between the sizes measured directly from the observed profiles and those measured from the best-fit Voigt or S\'{e}rsic profiles is $0.1 - 0.2$ dex, although there is considerable scatter. Finally, while we again caution that the top-hat profile fits are significantly worse for the majority of the objects (as seen in the bottom-left corner of Figure \ref{fig:ProfileCompare}), overall, these lower-limits to the NLR size are $0.3 - 0.4$ dex smaller than the sizes derived from Voigt or S\'{e}rsic fits.
	
	As a further test of the discrepancy between sizes measured directly from the observed spatial profile and those measured with S\'{e}rsic fits, we have re-measured the sizes of the obscured quasars from H13 directly from the observed Southern African Large Telescope (SALT) spatial profiles. The sizes that we measure are, on average, 0.2 dex larger than the sizes reported using a one dimensional S\'{e}rsic fit convolved with a Moffat profile, in agreement with the difference we report for the Gemini sample. We also re-fit the data from H13 using our two-dimensional procedure outlined above, and recover sizes for the objects that are within 0.1 dex of the sizes reported in H13. While H13 used a one-dimensional fitting procedure, the seeing was significantly larger ($\sim 2''$) than the size of the slit ($1.25''$) for their observations, so accounting for the slit width does not significantly affect the NLR size measurements. When we include the SALT data in fits to the size-luminosity relationship (Section \ref{sec:nlrsize_relationship}), we will use the two-dimensional Voigt and S\'{e}rsic fit size measurements for consistency with the Gemini observations and size measurements. 
	
	In \citet{liu2013}, the authors derive measurements of $R_{\mathrm{int}}$ using GMOS IFU data. Five of the objects in our sample of quasars overlap with the sample from \citet{liu2013}: J0210-1001, J0319-0019, J0319-0058, J0759+1339, and J0807+4946. As the \citeauthor{liu2013} sizes are not corrected for the seeing, we compare the sizes for these objects with the sizes we derive directly from the observed spatial profiles. Overall, the sizes agree to within 0.1 dex, and the differences are most likely due to the differences in the seeing between the observations. For an object where we have similar estimates for the seeing, J0210--1001, \citeauthor{liu2013} measure log($R_{\mathrm{int}}$/pc)$ = 4.2$, while we measure log($R_{\mathrm{int}}$/pc)$ = 4.1$. When we account for the seeing, however, we estimate a smaller size of only log($R_{\mathrm{int}}$/pc)$ \sim 3.7$, highlighting the importance of this correction. Another potential source of error on size measurements made from long-slit spectroscopy arises from the fact that the geometry of the NLR in our sample of quasars may not be perfectly round, and the sizes measured from an individual slit can suffer from projection effects. Evidence from \citet{liu2013} suggests that the hosts of these obscured quasars are round ellipticals, and H13 find that NLR sizes measured for the same objects at different position angles are very similar. Nonetheless, NLR projection effects could lead to sizes which are underestimated. 

\section{The Relationship between NLR size and AGN IR Luminosity}
\label{sec:nlrsize_relationship}
		
	We can use the size measurements made for our full sample of Type II quasars to better explore how the size of the NLR scales as a function of AGN luminosity. We plot the measured values of $R_{\mathrm{int}}$ for our sample on the NLR size vs. $L_{8 \mu \mathrm{m}}$ diagram from H13 in Figure \ref{fig:NLRsize}. In this figure, we use light blue squares to mark the objects from our sample, using the sizes derived from fits using a S\'{e}rsic profile. We also plot Type II Seyferts and quasars taken from the literature as described in H13. We plot the objects from \citet{greene2011} and \citet{liu2013} with downward facing arrows, as these measurements were made without accounting for the seeing. We also use colors to flag galaxies that may have a substantial contribution from star formation, as in H13. Using the criteria described in \citet{wright2010} and \citet{stern2012}, we mark objects with WISE color $[3.4]Ð[4.6] < 0.8$ with a light gray circle. We also mark those objects with $[3.4] - [4.6] < 0.6$ (a more relaxed demarcation) with dark gray circles. The objects in our sample span a range of $\log{(L_{8 \mu \mathrm{m}} / \mathrm{erg\, s}^{-1})} = 44.4 - 45.4$, and the measured sizes support the idea that for luminous Type II quasars there is a flattening of the relationship between NLR size and $L_{8 \mu \mathrm{m}}$ at quasar luminosities. The maximum NLR size and the luminosity at which the flattening is observed, however, depend on the method used to calculate NLR sizes.

	\begin{figure*}[ht]
	\epsscale{1.1} 
	\plotone{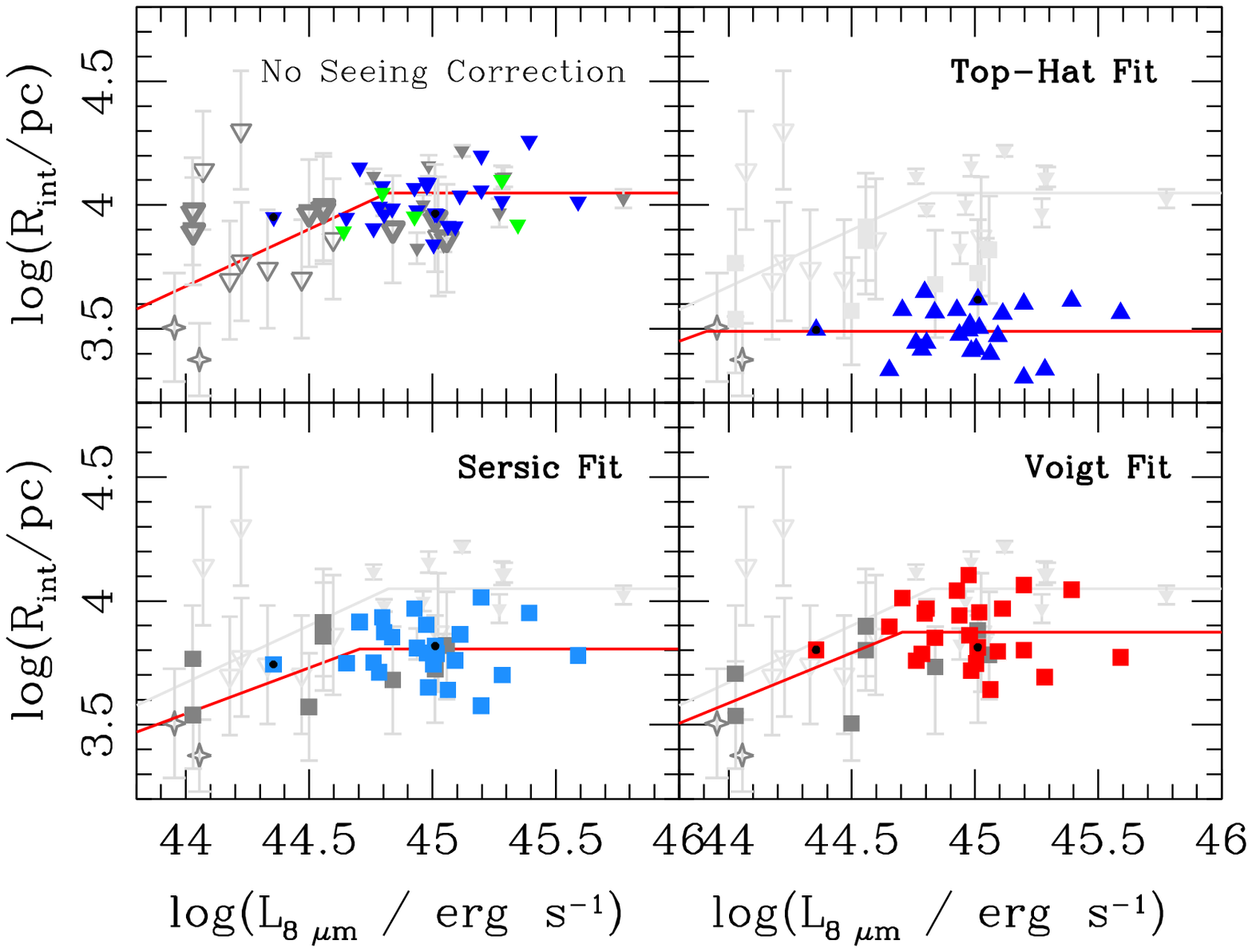}
	\caption{
	\label{fig:NLRsizemulti} Radii of the [OIII]$\lambda$5007-emitting region plotted against the AGN IR luminosity, focusing on the luminous end of the relation. The Type II quasars from our GMOS sample are given with colored symbols depending on the method of measuring $R_{\mathrm{int}}$. We plot sizes measured directly from the spatial profiles in the top-left corner using blue diamonds (for those objects where the spatial profile was not resolved, we plot an upper limit to the NLR size with a green square); we plot sizes measured using a top-hat fit in the top-right corner using blue triangles; we plot sizes measured using a S\'{e}rsic profile fit in the bottom-left corner using light blue squares; we plot sizes measured using a Voigt profile fit in the bottom-right corner using red squares. We also show fits to the data using both the objects in the current sample and objects from the literature, with symbols as in Figure \ref{fig:NLRsize}. In each portion of the plot, the objects included in the plot are given with dark grey points, while those excluded from the plot are given with light grey points. Finally, we plot the fit for the data without seeing corrections in each other section using a grey line. See the text for a description of the fitting procedure. We indicate the two radio-loud objects in our sample, J0843+2901 and J0943+0243, with black circles. Accounting for the seeing reduces the size at which the relationship flattens by $0.1 - 0.2$ dex.} 
	\epsscale{1.}
         \end{figure*}

	In order to examine the flattening of the relationship between NLR size and AGN luminosity, we fit the NLR size data following the method used in H13. The results of the fitting are presented in Figure \ref{fig:NLRsizemulti}, where we focus on the luminous end of the relationship. Our fitting method uses regression to fit the relationship piecewise: linear at low luminosities along with a function that flattens at some radius ($R_{\mathrm{0}}$) to stay at a ``maximum'' luminosity. Because the data from the literature was measured under different observing conditions, it is important to carefully choose which objects to include in our fits so as to include only objects where the sizes were measured in a similar way. Accordingly, in each portion of Figure \ref{fig:NLRsizemulti}, we show the Gemini sample with colored symbols, while the points included (excluded) in the fit are plotted with dark (light) grey symbols. While we include the lower luminosity points from \citet{fraquelli2003} and \citet{bennert2006} in order to constrain the luminosity at which the relationship flattens, we note that these sizes are also not corrected for the seeing, and as a result, we will not report the slopes to the best-fit relations.
	
	In the top-left corner of Figure \ref{fig:NLRsizemulti}, we plot the NLR sizes for objects in our Gemini sample measured directly from the observed spatial profile using blue diamonds (objects that are not resolved above the seeing are depicted by green triangles). For a correct comparison, we also plot the sizes we measured directly from the observed spatial profiles for the quasars observed in H13 as discussed in Section \ref{sec:nlrsizes}. For this fit, we calculate a value of $\log{(R_{\mathrm{0}}/\mathrm{pc})} = 4.1$ ($\sim 11$ kpc). In the top-right corner, we show the ``minimum'' sizes derived from the top-hat fits to the data from our sample, and measure $\log{(R_{\mathrm{0}}/\mathrm{pc})} = 3.5$ ($\sim 3$ kpc), which should be treated as a lower limit. In the bottom-left corner, we plot the sizes for the Gemini and H13 samples fit with a S\'{e}rsic profile, and measure $\log{(R_{\mathrm{0}}/\mathrm{pc})} = 3.8$ ($\sim 6.4$ kpc). Finally, in the bottom-right corner, we plot the sizes for the Gemini and H13 samples fit with a Voigt profile, and measure $\log{(R_{\mathrm{0}}/\mathrm{pc})} = 3.9$ ($\sim 7.4$ kpc). The resulting flattening of the relationship between NLR size and AGN luminosities implies that galaxy gas content, not the amount of ionizing photons from the AGN, is what limits the size of the NLR in the brightest quasars, supporting the results from H13. 

	\begin{figure*}[ht]
	\epsscale{1.1} 
	\plotone{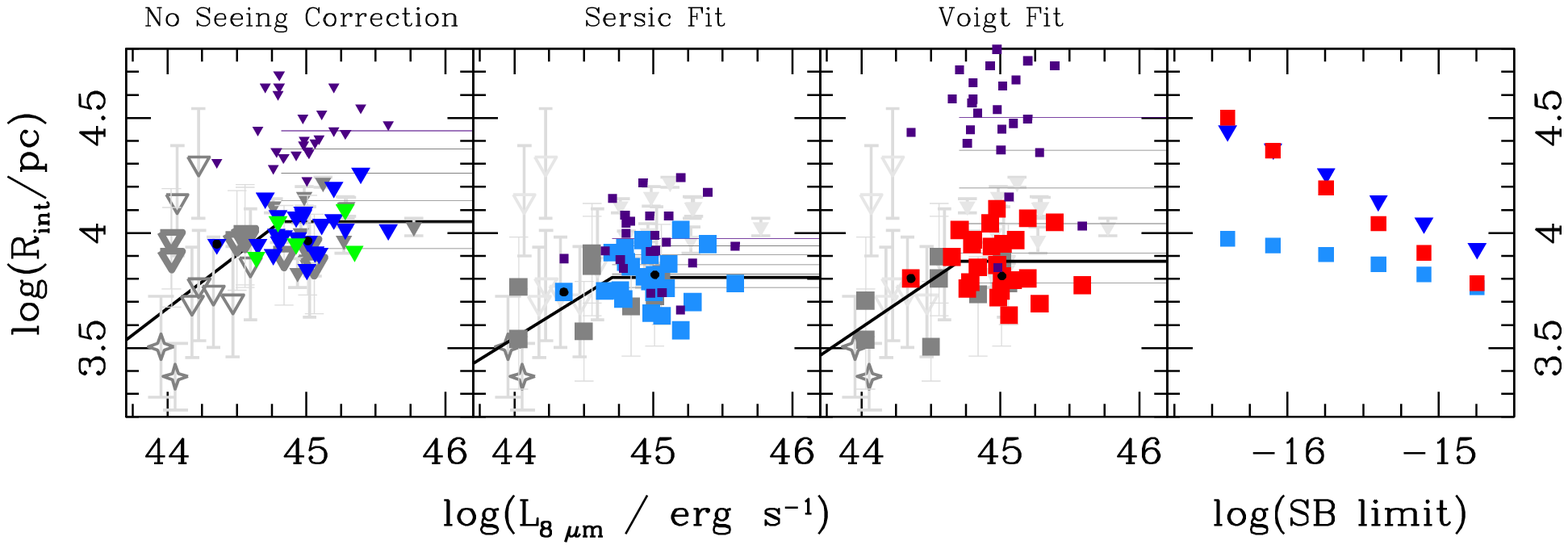}
	\caption{
	\label{fig:sb_explore} Radii of the [OIII]$\lambda$5007-emitting region plotted against the AGN IR luminosity, highlighting the effect of changing the surface brightness limit used to define $R_{\mathrm{int}}$. The points are the same as in Figure \ref{fig:NLRsizemulti}, and we additionally plot in purple our objects using sizes measured at a surface brightness limit of $0.04 \times 10^{-15}$ erg s$^{-1}$ cm$^{-2}$ arcsec$^{-2}$ (we also show the average for these points with a horizontal purple line). The horizontal grey lines indicate, from top to bottom, the average size estimated for the full sample using surface brightness limits of 0.08, 0.18, 0.4, 0.8, and $1.8 \times 10^{-15}$ erg s$^{-1}$ cm$^{-2}$ arcsec$^{-2}$. In the left panel, we plot the sizes measured directly from the observed spatial profiles, in the center-left panel, we plot sizes measured from fitting the observed profiles with a S\'{e}rsic profile, and in the center-right panel we plot the sizes estimated using a Voigt profile. On the far right, we plot these results but show the average sizes for each method as a function of the surface brightness limit (in units of erg s$^{-1}$ cm$^{-2}$ arcsec$^{-2}$), with the same points as in the left panels. Due to the extended wings on the Voigt profile, the sizes estimated are larger at fainter surface brightness limits than those estimated from a S\'{e}rsic profile. The sizes stay roughly constant across the entire range of luminosities, providing evidence that they are not dependent on our choice of limit for $R_{\mathrm{int}}$.}
	\epsscale{1.}
         \end{figure*}

	We note that the sizes we estimate for the NLR depend on the (arbitrary) surface brightness limit used to define $R_{\mathrm{int}}$. The limit of 10$^{-15}$ erg s$^{-1}$ cm$^{-2}$ arcsec$^{-2}$ for this analysis was chosen in order to compare to previous studies in the literature that probe AGN over a wide range in luminosity \citep[][H13]{greene2011,liu2013}, however it is instructive to explore how the NLR sizes depend on the adopted surface brightness limit. We therefore re-calculated the NLR sizes of our objects using the same procedure described above, but with $R_{\mathrm{int}}$ defined for cosmologically-corrected surface brightness limits that are both brighter and fainter than 10$^{-15}$ erg s$^{-1}$ cm$^{-2}$ arcsec$^{-2}$ (Figure \ref{fig:sb_explore}). We did not change our fits to the data, but merely evaluated the sizes at different surface brightness limits, for the observed profiles as well as the intrinsic S\'{e}rsic and Voigt profiles. To determine the minimum surface brightness that we can probe with our observations, for each of our objects we calculated the 3$\sigma$ limit for the continuum on either side of the [OIII] spatial profile, yielding an average minimum detectable surface brightness of $0.04 \times 10^{-15}$ erg s$^{-1}$ cm$^{-2}$ arcsec$^{-2}$. The sizes measured for this limit are shown by the purple points in Figure \ref{fig:sb_explore} (the average for these sizes is shown with a purple horizontal line), while the grey lines show the average sizes for the sample at surface brightness limits of 0.08, 0.18, 0.4, 0.8 and $1.8 \times 10^{-15}$ erg s$^{-1}$ cm$^{-2}$ arcsec$^{-2}$ (from the top line down). Figure \ref{fig:sb_explore} shows results for the observed, uncorrected profile (left), the S\'{e}rsic profile (center-left) and the Voigt profile (center-right). We note that this exercise is not applicable to the top-hat profile measurements, since the profile drops directly to zero at its outer extent and so the size is independent of the surface brightness limit. In the right panel, we show the average sizes measured as a function of surface brightness limit for all three methods simultaneously. 

These results indicate that regardless of the method used and the surface brightness limit adopted, the NLR sizes are still roughly constant with luminosity over the range probed by our sample. While the sizes estimated using a S\'{e}rsic profile are larger at fainter limiting surface brightness values, the effect is much stronger for those measured using a Voigt profile due to its more pronounced wings. At the lowest surface brightness limit used, we measure $\log{({R}/\mathrm{pc})} = 4.5$ (32 kpc) for the Voigt profile, and $\log{(R/\mathrm{pc})} = 4.0$ (10 kpc) for the S\'{e}rsic profile. The dispersion in the NLR sizes grows as we reach fainter surface brightness limits, from $\sim 0.1 - 0.2$ dex for the sizes measured using a Voigt profile, and $\sim 0.10 - 0.14$ dex for the sizes measured using a S\'{e}rsic profile, but the results are consistent with NLR sizes that do not depend on luminosity,  independent of our choice of surface brightness limit.

\section{Excitation Properties of the Outer Regions of Quasar Host Galaxies}
\label{sec:excitation}
	
	Figures \ref{fig:NLRsize} and \ref{fig:NLRsizemulti} demonstrate that, for luminous quasars, the size of the NLR (as measured by $R_{\mathrm{int}}$) is similar across a wide range of AGN IR luminosities. Recently, \citet{young2013} used Hubble Space Telescope (HST) narrow-band imaging to explore the size of the ionized regions for a sample of Type I quasars at $z \sim 0.1$. After careful quasar PSF subtraction, they find extended line-emitting regions in their sample between 0.5 and 5 kpc from the galaxy nucleus, in agreement with the extended NLR sizes that we observe in our full sample. While the luminosity range of the quasars in the \citet{young2013} sample (log$(L_{[\mathrm{OIII}]} / \mathrm{erg}\,\,\mathrm{s}^{-1}) = 41.0 - 42.6$) is an order of magnitude smaller than the luminosity range of the quasars targeted in our Gemini sample (log$(L_{[\mathrm{OIII}]} / \mathrm{erg}\,\,\mathrm{s}^{-1}) = 42.7 - 43.5$), these objects represent a useful comparison sample with observations that do not suffer from seeing effects. 
	
	Importantly, these authors present spatially resolved [OIII]$\lambda5007$ to H$\beta$ (as well as [OIII]$\lambda5007$ to [OII]$\lambda3723$) line ratios for their sample. Their results indicate that the primary source of ionization at large radii for the objects in their sample is star formation, in direct contrast to the results from \citet{greene2011} and \citet{liu2013}, who measure high ratios of [OIII]$\lambda5007$ / H$\beta$ across the face of the galaxies in their sample of Type II quasars, which indicates that ionization from an AGN is likely the cause of the observed [OIII] emission. This interesting discrepancy may be due to strong [OIII] emission from an AGN that originates near the nucleus and is being observed farther from the galaxy center due to seeing effects. If H$\beta$ is not similarly strong near the nucleus, as would be expected for an AGN, seeing effects would lead to artificially increased line ratios at large radii. 

	\begin{figure}[htbp]
	\epsscale{1.2} 
	\plotone{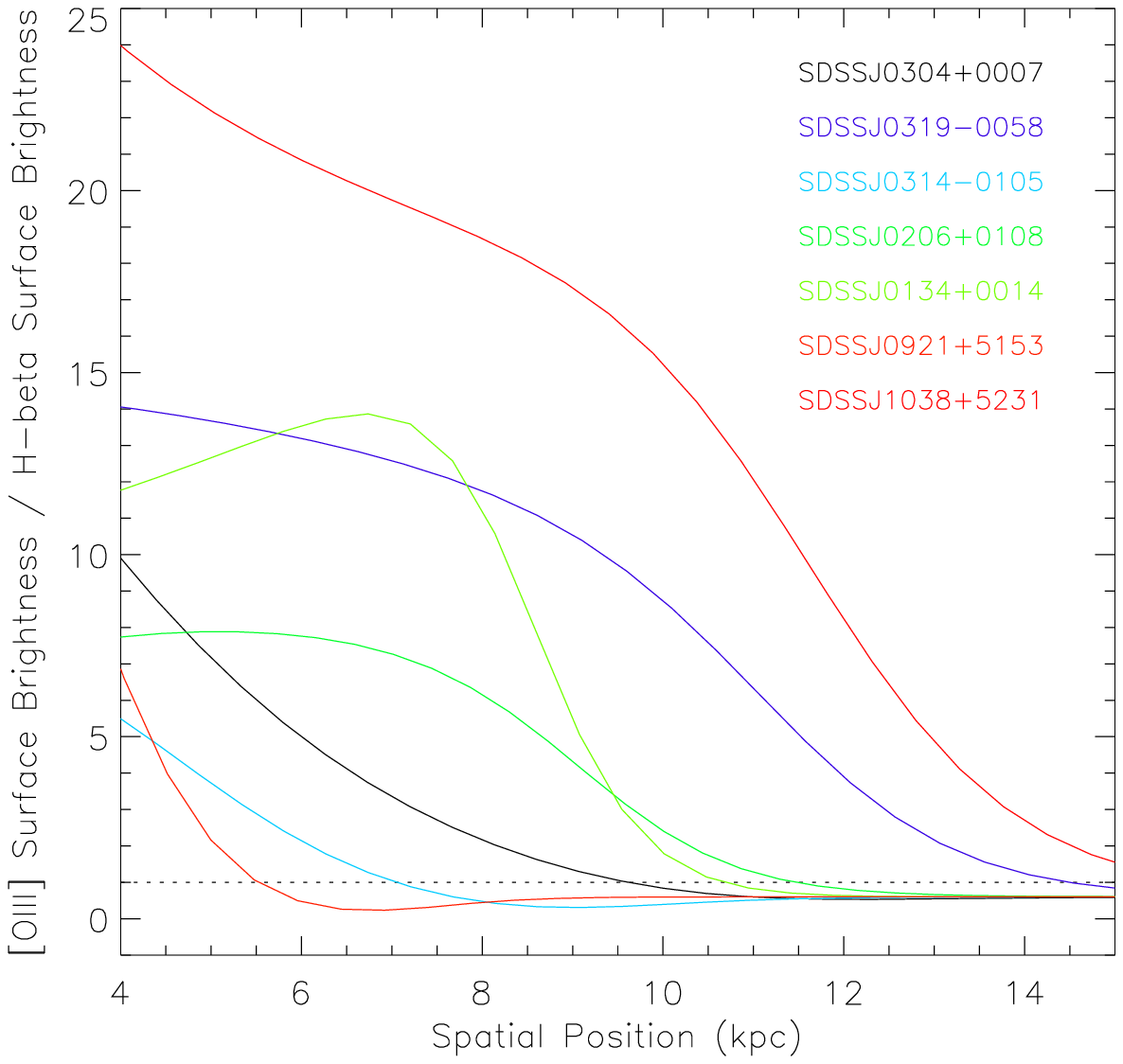}
	\caption{
	\label{fig:sbratiocomparison} Ratio of the intrinsic S\'{e}rsic surface brightness profiles of [OIII]$\lambda$5007 to H$\beta$ for objects with atmospheric seeing $< 0.55''$. These seven quasars show intrinsic [OIII] / H$\beta$ surface brightness ratios in excess of one (shown with a horizontal dotted line) out to large galactocentric radii, indicating that the mechanism for ionizing this emission is likely AGN activity rather than star-formation.} 
	\epsscale{1.}
         \end{figure}
	
	It is important to explore, then, the spatial extent of weaker emission features that originate primarily in star-forming regions. If star formation were the ionization source for the extended emission-line regions, lines such as H$\beta$ or [OII] would be observed to be strong compared to [OIII] at kpc scales. We examined the H$\beta$ emission feature for a sub-sample of objects in our Gemini sample with the lowest measure of the atmospheric seeing (we looked at the eight objects with seeing $< 0.55''$, although we excluded SDSSJ0319+0019, which does not show H$\beta$ in emission in both the Gemini spectrum as well as the SDSS spectrum), and fit the observed H$\beta$ spatial profiles in a similar manner to what was done for [OIII]. Using the intrinsic surface brightness profiles determined from the fits, we can take the ratio of [OIII] to H$\beta$ as a function of galactocentric radius, and show our results in Figure \ref{fig:sbratiocomparison}. As can be seen from the figure, [OIII] is much stronger than H$\beta$ even out at kpc scales, at which the picture of \citet{young2013} would predict strong H$\beta$ due to star formation. Our results indicate that AGN activity is likely the cause of the observed extended narrow-line emission for our sample of luminous Type II quasars. As an additional test, we fit the [OIII] and H$\beta$ profiles using a two component model: a central Dirac function to represent nuclear AGN emission, and then a Gaussian profile for extended emission. By then comparing the Gaussian fits for [OIII] and H$\beta$, we can explore the ionization properties of the gas in the absence of nuclear emission. Overall, while the fits were worse in this dual component fitting than the single component model fits described in Section \ref{sec:nlrsizes}, we find similar results to what was found using the intrinsic S\'{e}rsic profiles. 
	
	While further observations from space at high resolution should be performed on a full sample of Type II quasars to overcome potential seeing effects (similar to the analysis presented in \citet{young2013}), our results suggest that the large-scale NLR emission in our sample of sources is excited primarily by a quasar, and not star formation, which may be expected due to the larger AGN luminosities probed by our sample.

\section{Conclusions}
\label{sec:conclusions}

	We have performed Gemini long-slit spectroscopy  on a sample of 30 obscured luminous quasars to explore the proposed flattening of the relationship between NLR size and AGN luminosity first presented in H13. Our results indicate that, in luminous quasars, NLR size is roughly constant over an order of magnitude in AGN luminosity ($\log{(L_{8 \mu \mathrm{m}} / \mathrm{erg\, s}^{-1})} = 44.4 - 45.4$). Measurements of this size, however, are highly dependent on seeing effects, even in spatially resolved data. NLR sizes directly measured from observed spatial profiles overpredict those measured from profiles convolved with the seeing by 0.1 - 0.2 dex. As a result, while earlier data pointed to a limiting NLR size of $\sim 12$ kpc, our current results indicate that the size is approximately $6-8$ kpc. 

	When combined with other work, the fact that the Gemini sample is constant in NLR size across an order of magnitude in AGN luminosity (as shown in Figure \ref{fig:NLRsizemulti}) implies that NLR size in bright quasars is constrained by the availability of gas at the correct density and ionization state rather than the number of quasar ionizing photons. The exact limiting size is dependent on both the chosen model for the intrinsic surface brightness profile as well as the limiting surface brightness used to define $R_{\mathrm{int}}$, as shown in Figure \ref{fig:sb_explore}. The results presented in this paper support the claim made in H13 of a flattening of the slope of the NLR size -- AGN luminosity relation at the luminous end first put forward by \citet{netzer2004} and \citet{greene2011}. As discussed in H13, measuring the slope of the relationship is important to understand the geometry of the NLR for these objects, although these results indicate that perhaps correcting for seeing would lead to a steeper slope than what was given in H13. 

	Future research needs to be done targeting the most IR-luminous AGNs in order to trace the full extent of the turnover in the SDSS sample, ideally with IFU data as in \citet{liu2013}. Under excellent atmospheric conditions, IFU results would help to understand the true ionization source for NLR emission at large galactocentric radii. In order to overcome seeing effects, it would also be helpful to explore the NLR sizes for a sample of nearby luminous Type II AGNs using deep HST narrow-band imaging, in a manner similar to \citet{young2013}. 
		  
\acknowledgments 

We would like to thank the anonymous referee for their constructive comments which improved the final paper. KNH, RCH and ADM were partially supported by NASA through ADAP award NNX12AE38G and by the National Science Foundation through grant numbers 1211096 and 1211112. Support for the work of X.L. was provided by NASA through Hubble Fellowship grant number HST-HF-51307.01, awarded by the Space Telescope Science Institute, which is operated by the Association of Universities for Research in Astronomy, Inc., for NASA, under contract NAS 5-26555. This publication makes use of data products from the Wide-field Infrared Survey Explorer, which is a joint project of the University of California, Los Angeles, and the Jet Propulsion Laboratory/California Institute of Technology, funded by the National Aeronautics and Space Administration.




\end{document}